\documentclass[aps,prd,showpacs,nofootinbib,floats,floatfix,
preprintnumbers, 10pt]{revtex4-2} 
\usepackage{natbib}
\usepackage{comment}
\usepackage[utf8]{inputenc}
\usepackage{amsfonts,amsmath, mathrsfs}
\usepackage{graphicx}
\usepackage{amssymb}
\usepackage{bbm}
\usepackage{tensor}
\def\={\triangleq}

\def\lie{\pounds}
\begin{document}
\title{\textbf{Hawking radiation from  black holes in $2+1$ dimensions}}
\author{Akriti Garg}
\email{akritigarg571@gmail.com}
\affiliation{Department of Physics \& Astronomical Science, Central University of Himachal Pradesh, Dharamshala-
176215, India.}
\author{Ayan Chatterjee}
 \email{ayan.theory@gmail.com}
 \affiliation{Department of Physics \& Astronomical Science, Central University of Himachal Pradesh, Dharamshala-
 176215, India.}
\begin{abstract}
The paper develops a model to understand the effective quantum geometry of a black hole horizon and the emission of Hawking spectrum in $2+1$ dimensions. Using the algebra of Hamiltonian charges on the horizon, we establish that one should view the black hole horizon as formed out of quantised lengths of elementary quanta of value $8\pi \ell_{P}\, n$, where $n\in \mathbb{N}$, and $\ell_{P}$ is the Planck length. We determine the black hole entropy using this equidistant length spectrum in the microcanonical ensemble and show that its value is close to the Bekenstein-Hawking entropy. To evaluate the Hawking spectrum, we note that, to an observer near the black hole horizon, the entropy (or length of horizon cross- section) is related to the black hole energy. Hence, one may develop a formulation of length ensemble (similar to the area canonical ensemble of Krasnov) from which the black body spectrum may be obtained directly. This local observer perceives a Hawking spectrum whose temperature is modified by the Tolman factor.
\end{abstract}

\maketitle
\section*{}


The uniqueness theorems state that 
black holes in $4$- dimensions
may be characterised by their mass ($\mathscr{M}$),
charge ($\mathscr{Q}$) and angular momentum 
($\mathscr{J}$) only \cite{Hawking:1971vc, Carter:1971zc,Mazur:2000pn, Heusler:1996jaf}.
Due to this simplicity, black holes have a status similar to hydrogen atoms in 
the quantum theory of matter: they are considered as 
the first frontier to develop quantum geometry. 
This expectation arose due to several results during the last few decades. The foremost of them was 
the fundamental discovery
due to Bardeen, Carter and Hawking that black hole horizons in equilibrium, 
have laws of mechanics which are suspiciously similar to the laws of thermodynamics \cite{Bardeen:1973gs}. More precisely, 
they proved that (a) 
the surface gravity ($\kappa$) is a constant 
on the horizon, (b) area of the horizon cross- section $\mathscr{A}$ always increases in a classical physical process, and (c) horizons evolve to nearby equilibrium states in such a way that the horizon parameters 
are related through the following variational equation:
\begin{equation}\label{first_law}
\delta{\mathscr{M}}=(\kappa/8\pi )\,\delta\mathscr{A}+\Phi\,\delta\mathscr{Q}
+\Omega\,\delta\mathscr{J},
\end{equation}
where $\Phi$ is the electromagnetic potential, 
and $\Omega$ denotes angular velocity of the horizon\footnote{One may also state a certain version of 
the third law as well, although its ramifications
for gravity is not entirely clear \cite{Wald:1995yp, Wald:1997qp}}.  
This immediately suggested to Bekenstein that horizon
area should have the thermodynamic interpretation as
the entropy of black holes \cite{Bekenstein:1973ur, Bekenstein:1974ax}. He argued that
black hole horizons must be assigned an entropy equal to 
$\mathscr{S}=\mathscr{A}/4\ell_{P}^{2}$, 
where $\ell_{P}=\sqrt{\hslash}$ (in $4$ dimensions 
and with  $G=c=1$ units), is the Planck length. Once 
this point of view is accepted,
a look at the equation \eqref{first_law} tells us 
that horizons must be accorded a temperature $\kappa\ell_{P}/2\pi$. Hawking's semiclassical calculation
proved that large black holes ($\mathscr{A}\gg \ell_{P}^{2}$), produced from gravitational collapse of matter fields, do indeed behave like a perfect black body
and an observer at null infinity $\mathscr{I}^{+}$ receives a flux of radiation at $\kappa\ell_{P}/2\pi$ \cite{Hawking:1975vcx}.
One must note that the introduction of $\hslash$ in 
the first law, eqn. \eqref{first_law} is a clear indication that, at the fundamental level,
quantum effects control the geometry of horizons and 
even its evolution. \\

To formulate an effective quantum theory of
horizon, one faces an immediate question: is 
there a notion of Hamiltonian or energy associated with 
the horizon? Such a question is difficult to answer
from the eqn. \eqref{first_law}, since 
the equation has a curious mixture of 
global and local quantities: while
$\mathscr{A}$ is defined locally on the horizon, 
quantities like $\kappa$, $\mathscr{M}$, 
and $\mathscr{J}$ may only be defined with reference 
to observers at the asymptopia. Furthermore, in 
the quantum theory when Hawking radiation is in full swing,
and the horizon is evaporating, global definitions of 
horizon (event horizon for example), which forms the basis 
of eqn. \eqref{first_law}, may not be available. The isolated 
horizon (IH) formulation remedies this situation. One may 
show that the IH satisfies a \emph{quasilocal} 
version of eqn. \eqref{fig1}:
\begin{eqnarray}
    \delta{\mathscr{E}}=(\kappa_{IH}/8\pi )\,\delta\mathscr{A}_{IH}+\Phi_{IH}\,\delta
    \mathscr{Q}_{IH}
+\Omega_{IH}\,\delta\mathscr{J}_{IH},
\end{eqnarray}
where the quantities like $\kappa_{IH}$, $\mathscr{J}_{IH}$, $\cdots$ are all 
defined quasilocally on the horizon 
and $\mathscr{E}$ is the quasilocal energy (or 
the Hamiltonian function associated with 
the horizon generator) \cite{Ashtekar:2000hw, Ashtekar:2000sz, Ashtekar:2004cn, Chatterjee:2008if}. However, one 
still cannot integrate this equation to obtain 
the quasilocal energy since 
the exact dependence of the intensive quantities 
like $\kappa_{IH}, \Omega_{IH}$ are not known in terms
of the extensive variables 
$\mathscr{A}_{IH}, \mathscr{J}_{IH}$. The reason is the 
reliance only \emph{on} the horizon geometry, 
and one needs to go off the horizon to extract some 
reasonable features \cite{Ghosh:2011fc, Frodden:2011eb, Ghosh:2012wq, Ghosh:2013iwa}.\\

A few years back, Ghosh and Perez had argued that for 
$4$- dimensional stationary black holes in GR, one may
find a preferred class of observers near the horizon for whom, the first law of black hole mechanics reduces 
to a form $\delta \mathscr{E}=(\bar{\kappa}/8\pi )\delta A$ \cite{Ghosh:2011fc, Frodden:2011eb, Ghosh:2012wq}. Here, $\bar{\kappa}=1/\ell_{0}$ is the value of local surface gravity if the observer is at a proper distance $\ell_{0}$ from the horizon. This implies that
the local first law has a universal form for 
these preferred observers and is independent of the mass, charge or angular momentum of the black hole spacetime.
A simple integration of this equation leads to 
$\mathscr{E}=(\bar{\kappa}/8\pi ) A +o(\ell_{0})$ and so, for a large black hole and  $\ell_{0}\ll A$, 
the horizon energy as perceived by 
the local observer is 
$\mathscr{E}=(\bar{\kappa}/8\pi ) A$. 
This interpretation not only holds for 
the Kerr- Neumann class of solutions, 
but has been proved to be valid for isolated 
horizons (IH) as well \cite{Ghosh:2011fc, Frodden:2011eb, Ghosh:2012wq, Chatterjee:2015lwa, Tripathy:2023mpi}. The natural 
question is whether this local $\bar{\kappa}$ may 
admit a thermodynamical interpretation if 
one accounts for the quantum effects, or in 
other words, if one formulates a process through 
which the Hawking effect may be calculated for these 
local observers, would they have as a black body 
distribution at the temperature $\bar{\kappa}/2\pi$? 
Over the last few years, many \emph{plausible 
arguments} in favour of 
this question, in the affirmative, has been forwarded \cite{Ghosh:2011fc, Frodden:2011eb, Ghosh:2012wq, Tripathy:2023mpi}. 
Here, in this paper, we shall try to put this on 
strong footing in the context of a black hole 
horizon in $2+1$ dimensions. In particular, we 
shall show that an effective quantum process may be 
formulated for these $2+1$ horizons 
through which a black body emission spectrum with 
temperature $\bar{\kappa}/2\pi$ may be obtained. 
But why $2+1$ dimensions in particular?  \\

The gravitational field in $2+1$ spacetime 
dimensions has no propagating local degrees of 
freedom; the vacuum solutions imply that 
the spacetime is locally flat, or has a constant 
curvature depending on the cosmological constant \cite{Carlip:1998uc}.
However, the theory is not trivial since there 
does exist a black hole solution when
the cosmological constant is negative \cite{Banados:1992gq, Banados:1992wn}. Naturally, 
the discovery of this BTZ solution, named after 
their discoverers, is surprising since
black hole horizons are endowed with 
thermodynamical properties akin to perfect 
black bodies. More precisely, horizons have 
entropy ($\mathscr{S}$), 
and can emit Hawking radiation at a temperature 
proportional to their surface gravity $\kappa$. So,
horizons must admit (quite a large amount of) 
microscopic degrees of freedom to explain such 
thermal behaviour. However, the origin of such 
effective degrees of freedom remains unknown, 
although several well known proposals do exist. 
The foremost of them being that the 
states associated with the 
asymptotic conformal field theory at the BTZ 
boundary (the spacetime is asymptotically AdS3) give
rise to the statistical entropy of this horizon
\cite{Brown:1986nw, Strominger:1996sh, Strominger:1997eq, Carlip:1998wz, Carlip:1999cy, Dreyer:2001py, Kaundinya:2024wkb, Basu:2011qy, Afshar:2017okz}. This formalism has been successful 
to a great extent and has even been extended to 
obtain CFT, and hence black hole entropy, 
from geometrical structures near 
the horizon. Our proposal is to directly use the geometrical structure of horizon and extract effective 
microscopic degrees of freedom to explain 
black hole thermodynamics. \\

Our paper has two parts: the first part deals with 
classical horizon geometry, while the second part 
develops an effective quantum description of horizons.
To begin with, as our first objective, we shall show 
that 
for the BTZ black hole too, one may consider a 
special observer for whom 
the local first law, and horizon energy has the form 
proposed in \cite{Ghosh:2011fc, Frodden:2011eb, Ghosh:2012wq}. This is not surprising since the 
BTZ horizon is an isolated horizon \cite{Ashtekar:2002qc} 
and hence, is expected 
to be amiable to local observers and 
their thermodynamic interpretation.  More precisely,
we shall show that the universality of $\bar{\kappa}$
may be translated to imply the existence of a 
Gibbs- Euler equation for the energy $\mathscr{E}$ having the form  $\mathscr{E}=T\mathscr{S}$, with $T=\bar{\kappa}\ell_{P}/2\pi$, and an appropriate form for the entropy $\mathscr{S}$. \\

All these considerations are classical, and our next 
objective in this paper is 
to obtain an effective quantum description of 
this horizon from which the horizon entropy and 
the black body spectrum may be obtained. To that end,
we shall first show that one may view 
the horizon cross- section (which is a length), 
as formed out of geometrical length quanta which 
are integral multiples of $8\pi \ell_{P}$, which is similar to the Bekenstein- Mukhanov proposal 
for black holes in $4$- dimensional spacetimes. 
One may obtain the entropy of
these horizons in a microcanonical ensemble framework,
by counting the number of ways a classical area (length) 
may be obtained and using the Boltzmann formula, 
$\mathscr{S}=\ln \Omega$ (with $k_{B}=1$ units). We shall show 
that with appropriate choice of parameters, 
the value of $\mathscr{S}$ is quite close to  the Bekenstein- Hawking formula.  
Then, we use this discretised structure of 
the horizon to
obtain the Hawking radiation. Let us elaborate 
briefly on our assumptions in this regard. 
We shall view the process of Hawking radiation 
to be similar to an atomic transition process where 
an atom jumps from an higher excited state to 
a lower state with the emission of photons \cite{Krasnov:1996tb, Krasnov:1997yt}. 
Here, we shall reckon 
the black hole horizon to be a system being 
observed by the preferred local observer $\mathscr{O}$ placed at a proper distance $\ell_{0}$ from the horizon.
The horizon may similarly be thought of as 
making a transition from a state with
higher length (which is a state with higher entropy 
or higher local energy as viewed by $\mathscr{O}$) 
to a lower energy state by 
the emission of length quanta. Note that one 
should view the emission of these chuncks of length  
not merely as emission of gravitational degrees of 
freedom but also due to matter 
degrees of freedom as well, although we must admit that we
do not have a clear mechanism for such a process (where gravitational and matter degrees of freedom 
are exchanged/ related) and therefore, shall not 
elaborate on it here \cite{Perez:2023ctt, Barrau:2016qri, Barrau:2015ana}. Our system (which we shall always assume as being observed by $\mathscr{O}$) 
shall be a canonical ensemble in a 
thermal bath $\bar{\kappa}$. Since the energy of this
system is proportional to the length, we may also 
call this ensemble a length ensemble (similar to the area ensemble) \cite{Krasnov:1997yt}.
Using this ensemble and the appropriate partition function, we sum over all 
the states and determine the average length 
in the ensemble. We show that 
this transition process does indeed lead to an 
emission spectrum with the appropriate black body like 
spectrum at 
temperature $\bar{\kappa}/2\pi$, as required.\\

We proceed as follows: In the next section, Sec. \ref{sec1}, we recall 
the properties of the BTZ black hole and evaluate
the Newman-Penrose (NP) coefficients for 
this solution. The 
horizon of BTZ black hole is 
an example of isolated horizon and we shall use
this framework to extract 
the associated symmetries and Hamiltonian charges.
More precisely, we shall show in Sec.
\ref{sec2} that although 
the full spacetime is invariant 
under (apart from diffeomorphisms) 
local Lorentz group $SO(2,1)$, Lorentz boosts 
are residual symmetries of this horizon, and 
the horizon area (or length) is the corresponding 
Hamiltonian charge. Using the charge algebra, we 
shall argue that symmetry implies the existence of 
elementary length quanta on the horizon which 
have values proportional to $\ell_{P}$. The 
black hole entropy shall be evaluated by counting 
the number of states for a fixed classical length.
For this counting the microcanonical ensemble 
shall be used. In section 
\ref{sec3}, we shall begin to develop 
this model of determine the spectrum of Hawking flux of radiation emitted from 
the horizon in $2+1$ dimensions. Since the spectrum is perceived by the local observer $\mathscr{O}$, we show that
the temperature of the black body radiation is modified by the Tolman factor.
We discuss the ramifications and possible extensions of this 
result in the Discussion section.\\   


\section{The horizon of BTZ black hole}
\label{sec1}
Let us first discuss the geometry of 
the BTZ spacetime. The BTZ black hole
is a solution of $2+1$ gravity with negative 
cosmological constant \cite{Banados:1992gq, Banados:1992wn}. The metric in the Schwarzschild- like coordinates is:
\begin{equation}\label{btz_1}
    ds^{2}= -N^{2}\, dt^{2}+ N^{-2} dr^{2} + r^{2}\left[ d\phi +N^{\phi}\, dt\right]^{2},
\end{equation}
where the metric functions are defined as:
\begin{eqnarray}
N&=&\left[-8M+\frac{r^{2}}{l^{2}}+\frac{16 \,J^{2}}{r^{2}}\right]^{1/2}, ~N^{\phi}=\frac{-4\,J}{r^{2}},~\text{and } ~ l^{2}\Lambda=-1. 
\end{eqnarray}
For any given value of $\Lambda$, this is a 2- parameter family of metrics labelled by $M$, and $J$. The quantity $N$ is zero on the horizon, and 
has two roots, the inner horizon at $r_{-}$ and the outer horizon at $r_{+}$:
\begin{eqnarray}
r_{\mp}=4Ml^{2}\left[1\mp \left\{1-\left(\frac{J}{Ml}\right)^{2}\right\}^{1/2}\right].
\end{eqnarray}
Sometimes, it is better to use the 
advanced Eddington- Finkelstein coordinates. 
The BTZ black hole has the following metric 
\begin{equation}
    ds^{2}= -N^{2}\, dv^{2}+2dv dr + r^{2}\left[ d\phi +N^{\phi}\, dv\right].
\end{equation}
%
 Our black hole horizon of interest is at $r=r_{+}$, which has topology $\mathbb{S}^{1}\times \mathbb{R}$, and is generated by Killing vector, 
 $\ell^{a}=\partial_{v}-N^{\phi}(r_{+})\partial_{\phi}$.
 The expansion ($\theta_{\ell}$) along 
 the outgoing null horizon vector field $\ell^{a}$ is vanishing, denoted here by $\rho=0$, while 
  the expansion $\theta_{n}$ along the ingoing  null vector $n_{a}$ is negative, denoted here by $\mu<0$. So, the horizon at $r=r_{+}$, as we shall describe  below, is an example of a non- expanding horizon, and even more generally, a weakly isolated horizon \cite{Ashtekar:2000hw, Ashtekar:2002qc, Ashtekar:2000sz}.\\

To understand the geometry in full generality, one may consider the full spacetime metric  $g_{ab}=-2\,\ell_{(a}\,n_{b)}+m_{a}\,m_{b}$, of signature ($-++$). The triad null vectors ($\ell^{a}, n^{a}, m^{a}$), are such that $\ell\cdot n=-1, m\cdot m=1$, while all other dot products are zero. For the BTZ spacetime, these triads are given by:
\begin{eqnarray}
\ell^{a}&=&\left(\frac{\partial}{\partial v}\right)^{a}+\frac{N^{2}}{2}(\partial_{r})^{a}-N^{\phi}\, (\partial_{\phi})^{a},\\
n^{a}&=&-(\partial_{r})^{a} ,~~~~\text{and}~~~m^{a}=\frac{1}{r}(\partial_{\phi})^{a},
\end{eqnarray}
Note that on the horizon, where $N=0$, the horizon 
generator $\ell^{a}=\partial_{v}-N^{\phi}
(r_{+})\partial_{\phi}$. Using these triads, 
one may form the Newman- Penrose (NP) equations 
for the NP scalars which are \cite{Ashtekar:2002qc}:
\begin{eqnarray}
\nabla_{a}\ell_{b}&=&-\epsilon\, n_{a}\ell_{b}+\kappa_{NP}\, n_{a} m_{b}-\gamma\, \ell_{a}\ell_{b}+\tau\, \ell_{a} m_{b}+\alpha\, m_{a}\ell_{b}-\rho\, m_{a}m_{b},\label{gradell_eqn}\\
\nabla_{a}\, n_{b}&=&\epsilon\, n_{a}n_{b}-\pi\, n_{a}m_{b}+\gamma\, \ell_{a}n_{b}-\nu\, \ell_{a}\, m_{b}-\alpha\, m_{a}n_{b}+\mu\, m_{a}\, m_{b}, \label{gradn_eqn}\\
\nabla_{a}\, m_{b}&=&\kappa_{NP}\, n_{a}n_{b}-\pi\, n_{a}\ell_{b}+\tau\, \ell_{a}n_{b}-\nu\, \ell_{a}\ell_{b}-\rho\, m_{a}n_{b}+\mu\, m_{a}\ell_{b}
\label{gradm_eqn}.
\end{eqnarray}
%
These expansions are valid for the entire spacetime. For the BTZ solution, the expansion simplifies since the NP scalars have the following values:
\begin{eqnarray}\label{btz_values}
\epsilon&=&[(1/2)f^{\prime}(r)- r(N^{\phi})^{2}], ~ \gamma=0,~ \alpha=N^{\phi},~\kappa_{NP}=0, ~ \tau=N^{\phi}\\
\rho&=&(-1/2r)N^{2}, ~\pi=N^{\phi}, ~~ \nu=0, ~~ \mu=-(1/r).
\end{eqnarray}
Here, the quantity $\epsilon(r_{+})$ is the standard surface gravity, and from now on, shall denote by $\kappa$. These values shall be useful for determining the connection and horizon geometry in the next section.
Note that the BTZ horizon (denoted by $\Delta$), is 
an example of a weakly isolated horizon (WIH). The boundary conditions for the WIH are as follows \cite{Ashtekar:2002qc}:
 \begin{enumerate}\label{WIH}
     \item $\Delta=\mathbb{S}^{1}\times \mathbb{R}$.
     \item The horizon generator is expansion free $\theta_{(\ell)}=0$.
     \item All field equations are valid on $\Delta$, and the energy momentum tensor $T_{ab}$ satisfies dominant energy conditions with $ -T^{a}{}_{b}\,\ell^{b}$ being future directed 
     and null. .
     \item The connection on the normal bundle, also called the rotation one form ($\omega_{a}$), is Lie dragged along the horizon generator, $\lie_{\ell}\,\omega\,\=\,0$. 
     \end{enumerate}
One may determine the quantity $\omega_{a}$, from the equation \eqref{gradell_eqn}: $\nabla_{\underleftarrow{a}}\ell^{b}\,\=\, \omega_{a}\ell^{a}$, from which we get that (the symbol $\=$ shows that equality holds on $\Delta$ only and the arrow symbol $\alpha_{\underleftarrow{a}}$ implies that the index is pulled back to $\Delta$):
\begin{equation}
\omega_{a}\, \=\,-\kappa \,n_{a}+\alpha \, m_{a}.
\end{equation}
One may easily show from the above 
equation that the horizon generator 
$\ell^{a}$ is a Killing vector field on 
the horizon $\Delta$:
\begin{equation}
    \underleftarrow{\nabla_{(a}\,\ell_{b)}}
    \,\=\, 0\,.
\end{equation}
The fourth condition gives us 
the zeroth law $d\kappa=0$. The equation 
\eqref{gradm_eqn} shows that the
horizon cross- section does not evolve, 
and therefore
on $\Delta$, the horizon area 
element remains unchanged:
\begin{equation}\label{area_nchange}
    d\, m=0. 
\end{equation}
We are now going to 
use this broader geometric structure of WIH to develop 
our classical and effective quantum structure on 
the horizon.  
 
\section{Action, symmetries and Hamiltonian charges}
\label{sec2}
Given the boundary conditions for 
the black hole horizon specified through 
the WIH boundary conditions of the previous section, 
the next task is to construct 
the space of solutions of the theory of gravity which 
satisfy the WIH boundary conditions at the inner 
boundary, but is asymptotically AdS. Since our main 
concern shall be about the black hole horizon, we 
shall concentrate only at the WIH and assume that the 
quantities at the AdS boundary may be appropriately 
tamed. We also know that for GR in $2+1$ dimensions, 
one has the option of choosing various forms for 
action. The standard one is of course 
the Einstein- Hilbert action for 
metric variables \cite{Carlip:1998uc}.
One may also use a Palatini form of action, where 
the variables are $SO(2,1)$ lie algebra valued co-triads and connection \cite{Ashtekar:1989qc, Romano:1991up}. Another well known
theory is the Chern- Simons action based on $ISO(2)$ gauge group \cite{Achucarro:1986uwr, Witten:1988hc} \footnote{
The $2+1$ Palatini action based on any Lie group G may be shown to be equivalent, modulo surface terms, 
to the Chern- Simons theory based 
on the inhomogeneous Lie group IG \cite{Ashtekar:1989qc, Romano:1991up}.}. Here, we 
shall use the Palatini formalism.\\

The first order Palatini action is given by the following form \cite{Ashtekar:2002qc}:
\begin{eqnarray}\label{Palatini}
8\pi \, L\,[e, \, A]=e^{I}\wedge F_{I} -(\Lambda/6)\,\epsilon^{IJK}\,e_{I}\wedge e_{J}\wedge e_{K} \,  \, ,
\end{eqnarray}
where $e^{I}$ is the triad one- form, and $A^{I}$ is the $SO(2,1)$ Lie- algebra valued connection one-form.
The field strength corresponding to the connection $A^{I}$ is given by $F_{I}=dA_{I}+(1/2)\,\epsilon_{IJK}\, A^{J}\,\wedge \,A^{K}$.
One may determine the equations of motion using 
the variation of appropriate variables. For example,
the variation of connection $A^{I}$ leads to:
\begin{equation}\label{var_A}
    de^{I}+A^{I}{}_{J}\wedge e^{J}\equiv D\, e^{I}=0,
\end{equation}
where the connection $1$-form $A^{I}$ is 
expressed in terms of
$A_{IJ}\equiv -\epsilon_{IJK}\,A^{K}$
\footnote{This may be done since the adjoint 
representation of the Lie algebra of $SO(2,1)$ is with 
respect to the structure constants $\epsilon_{IJK}$. 
More precisely, given a Lie- algebra $\mathcal{L}$ 
with structure constants $C^{I}{}_{JK}$, the adjoint 
representation of $\mathcal{L}$ (by linear operators 
on $\mathcal{L}$) is defined by the mapping 
$\alpha^{I}\in \mathcal{L}\rightarrow (ad_{\alpha})_{I}
{}^{J}=\alpha^{K}\epsilon^{J}{}_{IK}$. Here, $A^{I}=-
(1/2)\,\epsilon^{IJK}\,A_{JK}$, and hence, 
$A^{K}\,\epsilon_{KIJ}=-A_{IJ}$.}. Here,
$D$ is the covariant derivative 
operator whose action on Lorentz vector $\lambda^{I}$
may be written as:
\begin{equation}
    D\,\lambda^{I}=d\lambda^{I}+A^{I}{}_{J}\, \lambda^{J}.
\end{equation}
The equation \eqref{var_A} implies that 
$A_{IJ}$ is a spin connection. A variation of 
the action with respect to the co- triad results in the Einstein equation:
\begin{equation}\label{var_e}
    F_{I}=(\Lambda/2)\,\epsilon_{IJK}\, e^{J}\wedge e^{K}.
\end{equation}
The epsilon tensor shall play an useful role and 
may be expressed in terms of the basis one forms:
\begin{equation}
    \epsilon_{IJK}=3!\,\ell_{[I}n_{J}m_{K]}.
\end{equation}
A simple calculation shows the following duality 
rules hold for the internal basis elements:
\begin{equation}
  \epsilon^{IJK}\, m_{K}=\epsilon^{IJK}\, 
\epsilon_{K}=\epsilon_{IJ}=\,2\,\ell_{[I}\, n_{J]}, ~~~\text{and} ~~
(1/2)\,\epsilon^{IJK}\,\epsilon_{IJ}=-m_{K}=-
\epsilon_{K} \, .
\end{equation}

The connection $A_{IJ}$ may be explicitly obtained 
for the WIH in terms of 
the Newman- Penrose scalars. First, note that
one may set a basis on the WIH $(\ell^{I}, n^{I}, m^{I})$ such that the action
of $\partial_{a}$ on them vanishes. Second,
since $\ell_{a}=\ell_{I}\, e^{I}_{a}$, we may 
rewrite: 
\begin{equation}\label{gradalb}
    \nabla_{a}\, \ell_{b}=e^{I}_{b}\, A_{aI}{^{J}}\, \ell_{J}.
\end{equation}
Similar expressions may be obtained 
for $n_{I}$ and $m_{I}$.
Using eqn. \eqref{gradell_eqn}, \eqref{gradn_eqn},
and \eqref{gradm_eqn}, the full $SO(2,1)$ connection in the spacetime may be determined from eqn.\eqref{gradalb} 
and is given by:
\begin{equation}\label{conn_1}
A_{aIJ}=- 2W_{a}\,\ell_{[I}n_{J]}+2N_{a}\,n_{[I}m_{J]} + 2U_{a}\,\ell_{[I}m_{J]}= W_{a} B_{IJ}+N_{a} P_{IJ} +U_{a} Q_{IJ}.
\end{equation}
This spacetime connection may also be rewritten in terms 
of the connection one form with dual basis: 
\begin{equation}\label{conn_2}
A_{aI}=-W_{a} m_{I}-N_{a}\, n_{I} +U_{a}\, \ell_{I},
\end{equation}
where the one forms in this connection coefficients are 
given in terms of 
the following NP scalars and basis one- forms:
\begin{eqnarray}
W &=&-\gamma \ell -\epsilon n+\alpha m \\
N&=&\tau \ell +\kappa_{NP}\, n -\rho m \\
U&=& -\nu \ell -\pi n +\mu m .
\end{eqnarray}
%
%
Note that $A_{I}$ simplifies on the horizon $\Delta$ since, as stated in 
the eqn. \eqref{btz_values}, 
\begin{equation}
    W\equiv\omega\=-\kappa\, n +\alpha \, m,~~, ~~ N\=0, ~~\text{and}~~ U\=-\pi \, n +\mu\, m .
\end{equation}
The $SO(2,1)$ symmetry of the bulk spacetime 
does not survive on the WIH, 
the boundary conditions are invariant under 
only a subset of this symmetry group. 
We now proceed to obtain the residual
symmetry on the BTZ horizon. To that end, first let 
us now study the transformation of 
this connection under the local Lorentz transformations in the full spacetime.
Under the action of local Lorentz 
transformations ($\Lambda_{IJ}$),
the basis changes are of three types \cite{Chandrasekhar:1985kt}, 
\begin{eqnarray}
\ell\rightarrow \xi \ell;\, n\rightarrow (1/\xi) n;\, m \rightarrow m,\\
\ell\rightarrow \ell;\, n\rightarrow n-c m +(1/2)c^{2} \ell ;\, m \rightarrow -c\ell +m,\\
\ell\rightarrow \ell -b m +(1/2)b^{2} n;\, n\rightarrow n;\, m \rightarrow -b n +m.
\end{eqnarray}
The corresponding transformation matrices have 
the following form (the $4$-dimensional system is discussed in \cite{Basu:2010hv, Chatterjee:2015lwa, Chatterjee:2020iuf}:
\begin{eqnarray}\label{Lor_trans_1}
\Lambda_{IJ}&=&-\xi \ell_{I}n_{J}-\xi\ell_{J}n_{I} +m_{I}m_{J}\, ,\label{tran1}\\
\Lambda_{IJ}&=&-\ell_{I}n_{J}-\{n_{I}-cm_{I}-(1/2)c^{2}\ell_{I}\}\,\ell_{J}+(m_{I}-c\ell_{I})\, m_{J}\, ,\label{tran2}\\
\Lambda_{IJ}&=&-n_{I}\ell_{J}-\{\ell_{I}-bm_{I}-(1/2)b^{2}n_{I}\}\,n_{J}+(m_{I}-bn_{I})\, m_{J}\label{tran3}.
\end{eqnarray}
The connection $A_{IJ}$ changes under these transformations in the following way:
$A_{IJ}\rightarrow A^{\prime}_{IJ}=(\Lambda^{-1})_{I}{}^{K}\,A_{KN}\,\Lambda^{N}{}_{J}+(\Lambda^{-1})_{I\,K}\, d\Lambda^{K}{}_{J}$.
Using the transformations in equations \eqref{tran1},
\eqref{tran2} and \eqref{tran3}, 
the changes in the connection components $A_{IJ}$
are obtained:
\begin{eqnarray}
W\rightarrow W-d\ln\xi ; U\rightarrow \xi U; N\rightarrow \xi^{-1} N\\
W\rightarrow W-cN ; U\rightarrow U-dc+cW-(1/2)c^{2}N; N\rightarrow  N\\
W\rightarrow W+bU ; U\rightarrow U; N\rightarrow  N +db-bW-(1/2)b^{2}U.
\end{eqnarray}
The first two equations show that transformations
map the WIH to the WIH. However, the third transformation is not a symmetry since
the transformation of $N$ is not homogeneous: whereas $N=0$ for black hole horizons, the transformation
maps it to a non- zero value. Hence, the residual symmetry on the WIH consists of a null boost in 
the $\ell- n$ basis, or equivalently, dual 
rotation in the $m$ plane; and null 
rotation in $\ell-m$ plane or a $n$ boost. 
The generators of all three transformations in eqns 
\eqref{tran1}- \eqref{tran3} are given by:
$B_{IJ}=2\,\ell_{[I}n_{J]}$, 
$P_{IJ}=2\,n_{[I}m_{J]}$ and 
$Q_{IJ}=\,2\,\ell_{[I}m_{J]}$. As argued above, 
on WIH,
the $P_{IJ}$ vanish. However, it shall 
be useful to note that these generators
satisfy the following algebra of $so(2,1)$:
\begin{equation}\label{alg_vec}
    [P, Q]_{IJ}=P^{I}{}_{K}\,Q^{K}{}_{J}-Q^{I}{}_{K}\,P^{K}{}_{J}=B_{IJ}, ~~ [B, Q]_{IJ}=-Q_{IJ}~~\text{and} ~~[B, P]_{IJ}=P_{IJ}.
\end{equation}
In the next sections we shall obtain 
the Hamiltonian charges associated with 
these residual symmetries.\\

\subsection*{The covariant phase space and Hamiltonian charges}
\begin{figure}[ht]
    \centering
\includegraphics[width=0.75\linewidth]{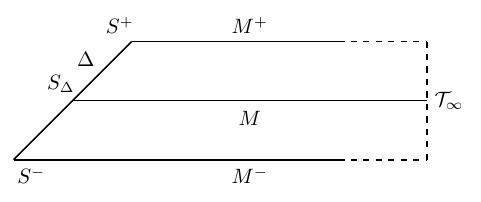}
    \caption{$M_{\pm}$ are  Cauchy slices enclose a region of space-time and intersect the $\Delta$ at $S_{\pm}$ respectively, and extend to the asymptotic infinity with the cylinder. Another Cauchy slice M is drawn which intersects $\Delta$ in $S_{\Delta}$}
    \label{fig1}
\end{figure}

Let us now determine the phase- space and 
the symplectic structure corresponding to 
the action given by eqn. \eqref{Palatini}. The 
space of solutions consist of 
all configurations of  
tetrad and spin- connection $[e, A]$, which 
solve the field equations,  
eqns. \eqref{var_A} and \eqref{var_e}, 
satisfy the WIH boundary conditions at $\Delta$, 
and have appropriate fall-offs at the asymptopia 
(see figure 1).  
Variation of the action \eqref{Palatini}, leads to 
a boundary term on- shell. This boundary
term is called the symplectic potential (a phase- 
space $1$- form but a $2$- form on the spacetime) 
$\delta L= \Theta(\delta)$, where
\begin{equation}
    \Theta(\delta)=e^{I}\wedge \delta A_{I}.
\end{equation}
The symplectic current, defined as: 
$J(\delta_{1}, \delta_{2})=\delta_{1}\Theta(\delta_{2})-\delta_{2}\Theta(\delta_{1})$, is a $2$- form on 
the phase- space
and a closed $2$- form on 
the spacetime, such that
$dJ(\delta_{1}, \delta_{2})=0$. 
Using $dJ=0$ over the region of 
the spacetime fig. \ref{fig1}, we get that:
\begin{equation}
    (\int_{M_{-}}-\int_{ M_{-}})\, J(\delta_{1}, \delta_{2})\, = \int_{\Delta}\, J(\delta_{1}, \delta_{2}) . 
\end{equation}
Here we have assumed that the integrals 
at infinity may be appropriately dealt with.
To construct a hypersurface independent symplectic 
structure, we need to show that on $\Delta$, $J(\delta_{1}, 
\delta_{2})\,=dj(\delta_{1}, \delta_{2})\,$, 
for some $j(\delta_{1}, \delta_{2})\,$.
On $\Delta$, the form of $J(\delta_{1}, \delta_{2})\,$
simplifies:
\begin{equation}\label{J_delta}
    J(\delta_{1}, \delta_{2})\,=\delta_{1}A^{I}\,\delta_{2}\, e_{I}-\delta_{2}A^{I}\, \delta_{1}\, e_{I} =-\delta_{1} \omega \,\delta_{2} m \,+ \, \delta_{2}\omega\, \delta_{1} m
\end{equation}
%
%
where we have used eqn. \eqref{conn_2}, and
$e^{I}_{a}=-n_{a}\,\ell^{I}-\ell_{a}\,n^{I}+m_{a}\,m^{I}$. However, we note that
one may define a potential $\psi$ for 
the surface gravity using
$\lie_{\ell}\,\psi^{\prime}=\ell\cdot\omega=\kappa$, such 
that $\psi^{\prime}=0$ at $S_{-}$. Using this potential,
the symplectic current on $\Delta$ may be 
rewritten:
\begin{equation}
    J(\delta_{1}, \delta_{2})\,=dj(\delta_{1}, \delta_{2})\,=-d[\delta_{1} \psi^{\prime} \,\delta_{2} m \,- \, \delta_{2}\psi^{\prime}\, \delta_{1} m],
\end{equation}
where we have used eqn. \eqref{area_nchange}.
The hypersurface independent symplectic structure is
then given by:
\begin{eqnarray}\label{sym}
8\pi \, \Omega(\delta_{1}, \delta_{2})=-\int_{\Delta}\left[\delta_{1}A^{I}\,\delta_{2}\, e_{I}-\delta_{2}A^{I}\, \delta_{1}\, e_{I} \right] -\oint_{S_{\Delta}}\left[\delta_{1}\psi^{\prime}\, \delta_{2} m -\delta_{2}\psi^{\prime}\, \delta_{1} m \right].
\end{eqnarray}
This symplectic structure has been 
used to prove the first law of black hole mechanics.
We shall jot down the main arguments, 
the details may be found in \cite{Ashtekar:2002qc}.
Let $t^{a}=\mathcal{B}_{\ell, t}\,\ell^{a}-
\Omega_{t}\,\phi^{a}$, be a vector field on $\Delta$.
Here $\mathcal{B}_{\ell, t}$ is some constant,
and $\phi^{a}$ is the angular Killing vector field
which describes rotational symmetry of WIH. The 
vector field $\phi^{a}$ must satisfy the following 
conditions: $\lie_{\phi}\,m=0$, $\lie_{\phi}\omega=0$,
and $\lie_{\ell}\,\phi=0$. Corresponding to $t^{a}$ 
there exists a phase space 
vector field $\delta_{t}\equiv\lie_{t}$ which is 
a Hamiltonian vector field. In particular, 
\begin{equation}
    \Omega(\delta, \delta_{t})=\delta H_{t},
\end{equation}
where 
$\delta H_{t}=\delta[E^{t}_{\infty}-E^{t}_{\Delta}]$ 
is a closed form on the phase space, while  
$E^{t}_{\infty}$ and $E^{t}_{\Delta}$ are 
the local energies defined at the asymptopia and
the WIH respectively. This implies that the
first law holds:
\begin{equation}\label{f_l}
    \delta E_{t}=\kappa_{t}\,\delta A +\Omega_{t}\, \delta J.
\end{equation}
Since the function $\delta H_{t}$ must be closed 
on the phase- space, $\kappa_{t}$ and $\Omega_{t}$
must be functions of horizon area and 
the angular momentum. This proves that 
the quasilocal first law of WIH is the necessary and sufficient condition for the vector field 
$\delta_{t}$ to be Hamiltonian.
\\

Are there other Hamiltonian charges associated with the WIH which may be obtained from 
the symplectic structure in eqn. \eqref{sym}? The 
answer is in the affirmative. Let us now determine 
the Hamiltonian charges associated with 
the local Lorentz transformations described in the 
equations \eqref{tran1}- \eqref{tran3}. 
The infinitesimal transformations associated
with the Lorentz matrices may be written as 
$\Lambda_{I}{}^{J}=(\delta_{I}{}^{J} +\epsilon_{I}
{}^{J})$, where $\epsilon_{IJ}$ is antisymmetric in 
the Lorentz indices,
lead to the following changes in the tetrad and 
connection one-forms:
\begin{eqnarray}
\delta_{\lambda} e^{I}&=&\epsilon^{I}{}_{J}\, e^{J},\\
-\delta_{\lambda} A^{IJ}&=&A^{IK}\,\epsilon_{K}{}_{J} 
+A^{JK}\,\epsilon_{I}{}_{K}+d\epsilon^{IJ}. 
\end{eqnarray}
Let us now determine if the vector fields 
corresponding to the symmetry of WIH in eqn.  
\eqref{tran1}- \eqref{tran2} are 
Hamiltonian. We use equation \eqref{sym}, and evaluate
$\Omega(\delta_{\lambda}, \,\delta)$. The boundary 
contribution to the symplectic structure vanish, and 
only the bulk part of eqn. \eqref{sym} contributes.
Note that the first term gives:
\begin{eqnarray}
\delta_{\lambda} A^{I}\wedge \delta_{2}\,e_{I}=-
(1/2)\epsilon^{IKL}\,\delta_{\lambda} A_{KL}\wedge 
\delta_{2}\,e_{I}
=d(\epsilon^{I}\delta e_{I})+\delta A_{K}\epsilon^{KJ}\wedge e_{J}.
\end{eqnarray}
The second term in the bulk symplectic 
structure is $-\delta 
A_{I}\epsilon^{IJ}\wedge e_{J}$ which cancels 
the second term in the above equation. So, 
the remaining contribution to 
the symplectic structure of eqn. \eqref{sym} is: 
$d(\epsilon^{I}\delta e_{I})$. Since this is a 
total derivative, contributions shall exist at 
the inner and the outer boundaries. If we assume 
appropriate fall- offs for the tetrads at 
the asymptotic infinity, we shall obtain only
an integral over the 
horizon cross- section at $S_{\Delta}$:
\begin{equation}
   8\pi \, \Omega(\delta_{\lambda},\delta )=\oint_{S_{\Delta}}\, \epsilon^{I}\, \delta \, e_{I}.
\end{equation}
For the transformation  $\delta_{\lambda}$ being a phase- space vector field corresponding to 
a Lorentz boost, with
$\epsilon_{IJ}=\ell_{[I}\, n_{J]}$,
the expression gives us: 
\begin{equation}\label{boost_l}
8\pi\, \Omega(\delta_{\lambda}, \delta)= -
\oint_{S_{\Delta}}\, \epsilon_{IJK}\delta e^{K} 
\times\epsilon^{IJ}=\oint_{S_{\Delta}}\delta\, m 
=\delta\, L \,,
\end{equation}
where $L$ denotes the length/circumference 
of the horizon cross- section. Therefore, 
our result implies that the area of 
horizon cross- section is a
Hamiltonian charge of the 
residual boost symmetry on WIH. 
This is not surprising since, as discussed 
earlier, the presence of boundaries often
give rise to global charges \cite{Chatterjee:2020iuf}.
Note that 
$\ell_{[I}n_{J]}=\epsilon_{IJ}$ is dual to 
$\epsilon_{K}=m_{K}$. The expression 
$\ell_{[I}n_{J]}=\epsilon_{IJ}$ is the generator of 
boosts, and the dual generates internal 
rotations of the circumference. So, the Hamiltonian 
charge corresponding to rotations is again 
the area of horizon cross- section:
\begin{equation}\label{area_rot}
8\pi\, \Omega(\delta_{\lambda}, \delta)= -\oint_{S_{\Delta}}\, \delta e_{I} \times\epsilon^{I}=-\oint_{S_{\Delta}}\delta\, m =-\delta\, L\, . 
\end{equation}
Now, since the generator the rotations 
is the angular momentum $J$, this charge must 
be proportional to the one obtained in \eqref{area_rot}. This gives, $8\pi \,\delta J\,=\delta L$ and hence, on integration,
the classical expressions of horizon area to 
be equivalent to the generator of rotations:
\begin{equation}\label{LJ_eqn}
    L=8\pi \, J.
\end{equation}
Other transformations give zero Hamiltonian charge. 
We now check the algebra of 
Hamiltonian charges to determine if 
they follow the algebra of vector fields.
For $\lambda_{1}$,  $\lambda_{2}$, and $\lambda_{3}$ being $B_{IJ}$, $P_{IJ}$,
and $Q_{IJ}$ respectively,
the expression is 
\begin{eqnarray}
 \Omega(\delta_{\lambda_{1}}, \delta_{\lambda_{2}})= 
 \{J, J_{1}\}=
J_{1}, ~~~
\Omega(\delta_{\lambda_{1}}, \delta_{\lambda_{1}})= 
\{J, J_{2}\}=
-J_{2}, ~~\text{and}~~\{J_{2}, J_{3}\}=J\,. 
\end{eqnarray}
This shows that the charges also satisfy 
the $SO(2,1)$ algebra, without any central charge.
We may now implement the equations on 
the phase space directly to 
the quantum state space by elevating
the Poisson brackets to commutator brackets.
On the space of quantum states, the $J_{3}$ charge
must be an integer multiple of $\hslash$:
$J\, | n\rangle= n\,\hslash \,|n \rangle$. 
Naturally, the eqn. \eqref{LJ_eqn}
may be written as:
 \begin{equation}\label{l_sp}
      L |n\rangle=8\pi \ell_{P} n \, |n\rangle.
 \end{equation}
This implies that length of 
the horizon cross- section is discretised with
an integral multiple of the Planck length. 
We shall call these effective 
quantum states, labelled by $n$, as \emph{punctures}, borrowing the nomenclature from loop quantum gravity.
Also, we shall assume that these states are distinguishable. An equispaced horizon area spectrum, as in eq. \eqref{l_sp}, 
resembling the Bekenstein- Mukhanov model has been 
proposed earlier for horizons in $2+1$ dimensions \cite{Birmingham:2001pj, Birmingham:2003wa}, as well as gravitational theories in for $3+1$ dimensions \cite{Dreyer:2002vy, Chatterjee:2020iuf, Devdutt:2026ljb}. 
Our calculation gives a simple illustration 
using the geometry of horizons directly.\\

One may now determine the entropy
be evaluating the number of ways to obtain
the given classical length $L$. For, 
$N$ a positive integer, the number of
combinations for a given $N=2^{N-1}$ and hence, 
the entropy is 
\begin{equation}\label{dis_entopy}
\mathscr{S}=\frac{L\ln 2}{8\pi \ell_{P}}=\frac{r_{+}\ln 2}{4 \ell_{P}}.
\end{equation}
The value of $\mathscr{S}$ is off by a factor of $(\ln 
2/2\pi)$ from the Bekenstein- Hawking result, but 
given the simple evaluation of discrete length 
eigenvalues, we should expect such deviations 
to occur.\\
%
\section{The Euler- Gibbs equation for local observers}\label{sec3}
Till now, we have restricted ourselves to geometry
on the horizon only. For example, our proof of the 
first law of black hole mechanics in eqn. \eqref{f_l} 
shows the variation of quantities which are 
defined \emph{on} the horizon. Similarly, the 
eigenstates and eigenvalues of the 
length operator in eqn. \eqref{l_sp} refers 
only to quantum nature of discreteness particular 
to the horizon $\Delta$. However, as 
argued in \cite{Ghosh:2011fc, Ghosh:2012wq, Ghosh:2013iwa}, we shall now try to 
demonstrate usefulness of the special class 
of observer to whom, the thermodynamics of 
black hole horizons is a natural map 
to usual thermal systems. 
Let us consider a stationary observer $\mathscr{O}$
located just outside the BTZ horizon $r_{+}$, 
at a proper distance $\ell_{0}$, such 
that $\ell_{0}\ll r_{+}$. We shall assume that the observer is like the zero angular
momentum observer (ZAMO), although they have
angular momentum $\sim o(\ell_{0})$. So, the observer
shall be at rest with respect to the horizon, and may 
be assumed to be have a preferred status for thermodynamics.
These observers follow the
Killing vector field of the BTZ black hole given by:
\begin{equation}
    \xi^{a}=\partial_{t}-N^{\phi}\, \partial_{\phi}.
\end{equation}
The four velocity of $\mathscr{O}$ is 
given by $u^{a}=\xi^{a}/||\xi||$.
One may determine the proper distance $\ell_{0}$ from 
the metric eqn. \eqref{btz_1}:
\begin{equation}
    \ell_{0}=\int_{r_{+}}^{r}\frac{dr}{N}=\ell 
    \ln\left[1+\sqrt{\frac{r^{2}- r_{+}^{2}}{r^{2}-
    r_{-}^{2}}} \right].
\end{equation}
The local surface gravity is 
$\bar{\kappa}=\kappa/||\xi||$,
and the energy $\mathscr{E}=T\mathscr{S}$ \cite{Ghosh:2011fc, Ghosh:2012wq, Ghosh:2013iwa, Tripathy:2023mpi}, with 
$T=\ell_{P}\bar{\kappa}/2\pi$. To this observer,
the system would behave as a thermal system with energy $\mathscr{E}$ and having thermal emission at
the black body temperature $T$.\\

We shall now argue that the local first law $\mathscr{E}=(\bar{\kappa}/8\pi )\, \mathscr{A}$, is
a natural consequence of the near horizon metric
as perceived by the observer. Note first that
the near horizon metric for the BTZ black 
hole is given by \cite{Carlip:1999cy}:
\begin{equation}
    ds^{2}=-r^{2}\, d\tau^{2}+\frac{dr^{2}}{\kappa^{2}}+2r^{2}N(\theta)d\tau\, d\theta
    +2rM(\theta)dr\, d\theta +X(\theta)\, d\theta^{2}
    + \cdots \, ,
\end{equation}
where $N(\theta), M(\theta)$ are
functions of spacetime. The natural triads 
associated with this metric are:
\begin{equation}
    e^{0}=r\, d\tau + \cdots , ~~\text{and}~~e^{1}=\frac{dr}{\kappa} +\cdots \, .
\end{equation}
Here, $\partial_{\tau}$ is the boost
Killing vector field generating the horizon $\Delta$ (this horizon is Rindler like for $\mathscr{O}$).
The observer moving with $4$- velocity 
$u^{a}=(1/r)\partial_{\tau}$, transforms and mixes the vector fields among themselves:
\begin{eqnarray}
u^{a}\nabla_{a}\, e^{b}_{0}&=&(\kappa/r)\, e_{1}^{b} + o(r)\, e_{0}^{b} + o(r) \cdots ,\\ 
u^{a}\nabla_{a}\, e^{b}_{1}&=&(\kappa/r)\, e_{0}^{b} + o(r)\, + \cdots .
\end{eqnarray}
For such observers which are at a proper 
length $\ell_{0}$ from the horizon, there does exist 
a preferred triads which are parallely 
transported along $u^{a}$, and are given by:
\begin{eqnarray}
    e^{0\,\prime}&= & e^{0}\, \cosh (\tau/\ell_{0}) +e^{1}\, \sinh (\tau/\ell_{0})\\
     e^{0\, \prime}&= & e^{0}\, \sinh (\tau/\ell_{0}) +e^{1}\, \cosh (\tau/\ell_{0}).
\end{eqnarray}
These tetrad frames are lie dragged by the Killing vectors in the following way:
\begin{equation}
    \lie_{\tau}\, e^{0\, \prime}=(1/\ell_{0})\, e^{1\,\prime}, ~~\text{and} \,~~  \lie_{\tau}\, e^{1\,\prime}=(1/\ell_{0})\, e^{0\,\prime}.
\end{equation}
Such a transformation may be viewed as
the the following derivative action:
\begin{equation}
    \lie_{\tau}\, e^{I}=\epsilon^{I}{}_{J}\, e^{J},
\end{equation}
with $\epsilon^{1}{_{0}}=1$, while $\epsilon^{0}
{}_{1}=1$, and the boost parameter is $1/\ell_{0}$. More precisely,
if $\Lambda_{IJ}=(\delta_{IJ} +\rho \,\epsilon_{IJ})$, then $\rho=1/\ell_{0}$. 
This should be viewed as a Lorentz- Lie 
derivative such that the action of diffeomorphisms 
(which shall induce transformations on 
the phase- space) is to be understood 
as a local Lorentz transformation \cite{Ortin:2002qb, Chatterjee:2015lwa}.
We have already noted in the previous section, eqn. \eqref{boost_l}, that
local Lorentz boost $\epsilon_{IJ}=\rho \,\ell_{[I}\, n_{J]}$ must give $L\, \rho/8\pi$ as the Hamiltonian charge. At the same time,
the boost generator for the observer $ \mathscr{O}$ near $\Delta$ is similar to generator of Hamiltonian or energy $\mathscr{E}$ \cite{Padmanabhan:2010zzb, Chatterjee:2015lwa}. So, on equating, we get:
\begin{equation}\label{eg_form}
    \mathscr{E}=L/(8\pi \ell_{0})=\bar{\kappa}L/(8\pi)\, .
\end{equation}
This is the desired result where the energy
as perceived by the local observer is in the Euler- Gibbs form. However, we must point out that the 
value of entropy which one may derive from the 
eqn. \eqref{eg_form} does not match with the 
value obtained from eqn. \eqref{dis_entopy}, although we shall use this discretised form in the following 
section.\\
\section{Length ensemble and Hawking spectrum}\label{sec4}
We are now in a position to answer the question 
related to the Hawking spectrum. The quantum states 
of the system are given in eqn. \eqref{l_sp} which 
are defined as the punctures and form the length of 
horizon cross- section. Note that the system 
is assumed to exchange energy (punctures) with 
the surroundings, and hence we shall need to 
construct the appropriate description for 
determining the probability of the thermal state.
Let us assume that the probability of finding the 
black hole horizon in a particular state $\psi$ is:
$p(\psi)\sim \exp [-\gamma\, L(\psi)]$, where 
$\gamma$ is a real parameter playing the role
of canonical conjugate to the horizon length.
Using this probability distribution, we construct the 
following \emph{partition} function \cite{Krasnov:1996tb, Krasnov:1997yt}:
\begin{equation}
Z(\gamma)=\sum_{\psi}\,\exp{-\gamma\, L(\psi) }.
\end{equation}
Using the expression of the length spectrum in eqn. \eqref{l_sp},
this sum over states may carried out to obtain
\begin{equation}\label{z_prod}
    Z(\gamma)=\prod_{n}\frac{1}{(1-e^{-\gamma \, L_{n}})},
\end{equation}
where the product is over the punctures/ tiles 
denoted by $n$. Therefore, it follows that
the average length is:
\begin{equation}\label{len_th}
    L(\gamma)=\sum_{\psi} L(\psi)\, p(\psi)=-\frac{d}{d\gamma}\ln Z =\sum_{n}\, L_{n}\,\, \frac{1}{(e^{\gamma L_{n}}-1)}\, ,
\end{equation}
where $p(\psi)=\exp[-\gamma \, L(\psi)]/Z$ and we have used the eqn. \eqref{z_prod}. 
\\

To determine the thermal spectrum,
we shall use the model of an atom making transitions through spontaneous and stimulated emissions and absorption processes.
Let $A_{\omega}$ be the coefficient of spontaneous 
emission for the mode $\omega$. We shall also take it 
to define the mean population $\mathcal{N}_{\omega}$
for this mode. Let us also denote 
the coefficient of stimulated emission 
$B^{12}_{\omega}$. So, if  
$\mathcal{N^{\prime}}_{\omega}$ denotes the
number of incoming quanta, the population of this 
mode in the outgoing radiation 
due to stimulated emission is 
$B^{12}_{\omega}\,\mathcal{N^{\prime}}_{\omega}$. 
Similarly, if the coefficient of absorption is 
$B^{21}_{\omega}$, and 
$\mathcal{N^{\prime}}_{\omega}$ is
the number of incoming quanta, 
the population of this mode due to absorption is 
$B^{12}_{\omega}\mathcal{N}_{\omega}$. We now determine the relation between the coefficients
$B^{12}_{\omega}$ and $B^{21}_{\omega}$.
Let $W_{\psi\rightarrow \psi^{\prime}}$ be the transition rate to go from the state ${\psi\rightarrow \psi^{\prime}}$, and that $p(\psi)$ is the probability of  horizon to be in state $\psi$.
\begin{figure}[t]
\begin{center}
\includegraphics[width=0.75\linewidth]{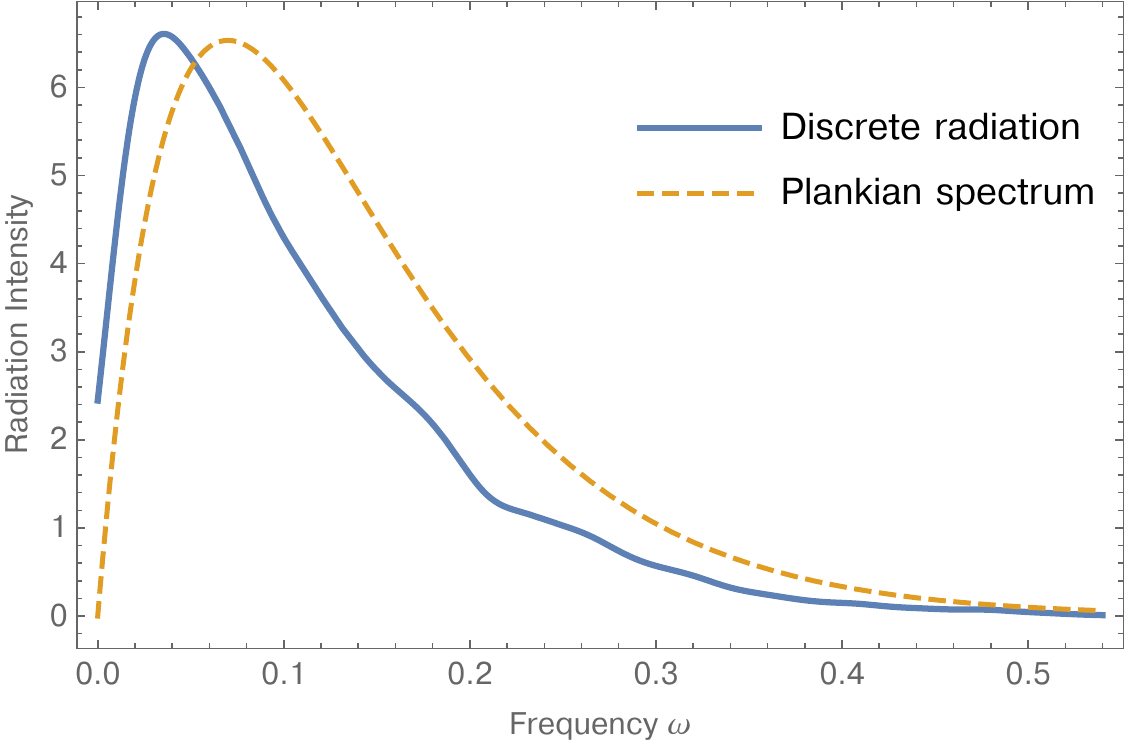}
\caption{The blue curve in the figure 
shows (a smoothened version of) number of punctures or length tiles emitted, as simulated through a Monte Carlo method. On the other hand,
the orange curve in the figure shows 
the plot of intensity as obtained from the spectrum 
eqn. \eqref{hawk_spec}. The matching of 
the two graphs is good in the 
large number of code runs.}
\label{fig_2}
\end{center}
\end{figure}
Then we get 
the following ratio of $B$- coefficients \cite{Krasnov:1997yt}:
\begin{equation}\label{b_ratio}
\frac{B^{12}_{\omega}}
{B^{21}_{\omega}}=\frac{\sum_{\psi,\, 
\psi^{\prime}}\,  \, p(\psi^{\prime})\, 
W_{\psi^{\prime}\,\rightarrow \psi}}{\sum_{\psi 
\psi^{\prime}}\, p(\psi)\, W_{\psi\,\rightarrow 
\,\psi^{\prime}}}.
\end{equation}
However, since the two states of the black hole horizon have different lengths, their probabilities 
must be different, dependent on their entropies. 
Using the value of discretised entropy eqn. 
\eqref{dis_entopy}, we get:
\begin{equation}
    p(\psi)=e^{\gamma (L-L^{\prime})}\, p(\psi^{\prime})=e^{\gamma\mathscr{E}16\pi^{2}/(\bar{\kappa} \ln 2)}\, p(\psi^{\prime})
=e^{\beta \hslash\omega} 
\, p(\psi^{\prime})\, ,
\end{equation}
where we have used $8\pi\gamma/(\ln 2)=\beta=2\pi/\bar{\kappa}$. Since the transition rates $W$ must be same for
transitions to and from the states, the eqn. \eqref{b_ratio} gives: $B^{21}_{\omega}= e^{\beta 
\hslash\omega}B^{12}_{\omega}$.
Define the absorption probability:
\begin{equation}
    \Gamma_{\omega}=B^{21}_{\omega}-B^{12}_{\omega}=(e^{\beta\,\hslash \omega} -1)\,B^{12}_{\omega}.
\end{equation}
Now since by the transition rules, $A_{\omega}=B^{12}_{\omega}$, this also gives 
the black body spectrum at the temperature
$\bar{\kappa}/2\pi$:
\begin{equation}\label{hawk_spec}
    A_{\omega}=n_{\omega}=\frac{\Gamma_{\omega}}
{e^{\beta\,\hslash\, \omega}-1}\, .
\end{equation}
\\
To see how the smooth Hawking spectrum of eqn. 
\eqref{hawk_spec} matches to the 
discretised emission of length/tiles or punctures
from the horizon, we simulate the process through
a random Monte Carlo method based on the length
spectrum in eqn. \eqref{l_sp}. We may observe from the fig. \ref{fig_2} that the blue line, which is a smoothened version of the discretised emission, matches quite closely to the Hawking spectrum.\\

\section{Discussion}
The paper has several results which we 
recall before we discuss their
ramifications. First, we have shown that
a black hole horizon, modeled 
after a WIH in $2+1$ dimensions breaks 
the $SO(2,1)$ local Lorentz symmetry of 
the spacetime and retains only a part which includes
a boost in the null $\ell-n$ plane, or a rotation 
along the spacelike $m$ vector. This result is 
not surprising since in $4$- dimensions too, 
the WIH reduces the $SL(2, \mathbb{C})$
to $ISO(2)\ltimes \mathbb{R}$ \cite{Basu:2010hv, Chatterjee:2020iuf}. Second, we 
show that generators of 
the residual algebra gives rise to genuine charges
on the horizon, and in particular the boost generator
is a Hamiltonian vector field for the horizon area.
We use this argument to show that one may 
construct a quantum state space \emph{on} the
horizon which give rise to equidistant 
length quanta, eqn. \eqref{l_sp}. Thirdly,
we show that the Euler- Gibbs equation 
$\mathscr{E}=T\mathscr{S}$, is generically valid for 
observers near the horizon. Fourth, we use the 
discretised length quanta to model the Hawking 
radiation process, and show that Hawking spectrum is 
obtained naturally through this 
atomic transition-like model. \\

Let us now discuss these results one by one.
We first note that breaking of gauge symmetries
due to boundary conditions is 
natural. In particular, in GR, it is well known that
at the boundaries, not all diffeomorphisms are
preserved, and the ones which do preserve them 
manage to generate physical degrees of freedom
associated with that boundary: for example, 
mass of spacetime 
may be viewed as a global charge due to such 
broken symmetries. Here also we have 
local $SO(2,1)$ Lorentz symmetries in the full 
spacetime, and the presence of WIH does lead to
field configurations where not all of
these symmetries are preserved. However, just 
like diffeomorphisms, here too, 
the Lorentz boost $B_{IJ}$
on the horizon generates the horizon area in 
the sense that the area of cross- section is 
the Hamiltonian charge. However, not all generators 
lead to non- zero charge; this is not unexpected 
since the WIH is a non- expanding horizon, and 
the vanishing of horizon area growth 
leads to vanishing charges for the 
generators $P_{IJ}$ and $Q_{IJ}$.
We then use this expression of classical area 
to construct an operator which would act 
on the quantum states residing on $\Delta$. 
These quantum states are obtained through the 
algebra of residual symmetries. We have obtained a
length operator which has 
eigenvalues $8\pi\ell_{P}\, n$, where 
$n$ is positive integer ($\mathbb{N}$). 
Such an equidistant spectrum for the area 
operator of the Bekenstein- Mukhanov type
model has been proposed earlier as well based on 
several theoretical as well as analysis
of quasinormal modes on the horizon \cite{Birmingham:2001pj, Birmingham:2003wa}. One may also
obtain an eigenvalue of $8\pi\ell_{P}\sqrt{j(j+1)}$
as in the loop quantum gravity models, although
the states for this computation are spread throughout 
the spacetime, and not merely confined to the horizon 
$\Delta$ \cite{Frodden:2012nu}. 
Our model is comparatively simpler and the 
states are attached to the horizon only.   
The black hole entropy in eqn. \eqref{dis_entopy} 
obtained through the counting of horizon states do
lead to an expression which is close to 
the Bekenstein- Hawking value. \\

To obtain the Hawking spectrum, we introduce a 
canonical length ensemble as in \cite{Krasnov:1997yt}. Here,
to create that ensemble, we have shown, using 
the $2+1$ formulation of WIH, that 
such an ensemble may be obtained for an observer
near the horizon $\Delta$. Only for those 
observers, a well defined notion of 
energy proportional to the horizon length 
does exist \cite{Ghosh:2012wq, Ghosh:2013iwa}.
Using this system, we show that the system is 
thermal and has Hawking spectrum at the temperature
$\bar{\kappa}$ which is a correction of the
temperature at 
the asymptotic infinity with the Tolman factor. Generalisation 
of this method to black holes in $4$- dimensions 
shall be carried out in the future.\\

\begin{acknowledgments}
AG acknowledges the financial support in the form of fellowship from Central University of Himachal Pradesh. The authors thank Theory Division, SINP Kolkata and for an academic visit. We also thank Prof. Amit Ghosh for encouragement and his insights during discussions.   
\end{acknowledgments}

\bibliographystyle{apsrev4-2} 
\bibliography{bibl}

\end{document}